\documentclass[apj]{emulateapj}
\usepackage{lscape}
\usepackage{apjfonts}
\usepackage{graphicx}
\newcommand\teff{T$_{\rm eff}$}
\newcommand\logg{log $g$}

\begin{document}
\title{The Effect of Differential Limb Magnification on Abundance Analysis of
Microlensed Dwarf Stars}
\author{
Jennifer A.~Johnson, Subo Dong\altaffilmark{1} and Andrew Gould}
\affil
{Department of Astronomy, Ohio State University,
140 W.\ 18th Ave., Columbus, OH 43210, USA 
}
\altaffiltext{1}{Current address: Institute for Advanced Study, 1 Einstein
Drive, Princeton, NJ 08540, USA}
\email{jaj,dong,gould@astronomy.ohio-state.edu}

\begin{abstract}

Finite source effects can be important in observations of gravitational
microlensing of stars. Near caustic crossings, for example, 
some parts of the source
star will be more highly magnified than other parts. The spectrum
of the star is then 
no longer the same as when it is unmagnified, and measurements
of the atmospheric parameters and abundances will be affected. The 
accuracy of abundances measured from spectra taken during
microlensing events has become important
recently because of the use of highly magnified dwarf stars to probe
abundance ratios and the abundance distribution in the Galactic bulge. While the
abundance ratios in general agree with the giants, with the possible
exception of [Na/Fe], there may be more dwarfs than giants with
high [Fe/H].
In this paper, we investigate the effect of finite source effects on
spectra by using magnification profiles motivated by two events to
synthesize spectra for dwarfs between 5000K to 6200K at solar
metallicity.  We adopt the usual techniques for analyzing the
microlensed dwarfs, namely, spectroscopic determination of
temperature, gravity, and microturbulent velocity, relying on
equivalent widths. We find that ignoring the finite source effects for
the more extreme case results in errors in \teff $< 45$K, in \logg{}
of $<$0.1 dex and in $\xi$ of $<$0.1 km/s. In total, changes in
equivalent widths lead to small changes in atmospheric parameters and
changes in abundances of $<$0.06 dex, with changes in [\ion{Fe}{1}/H]
of $<$0.03 dex. For the case with a larger source-lens separation, 
the error in [\ion{Fe}{1}/H] is $< 0.01$ dex. This latter
case represents the maximum effect seen in events 
whose lightcurves are consistent with a point-source
lens, which includes the majority of microlensed bulge dwarfs published
so far.

\end{abstract}

\section{Introduction
\label{sec:intro}}

A microlensing event occurs when a stellar mass object passes between the
observer and a background source. For events in the Local Group, the
source is a star. If the source is finite, rather than infinitely 
small, parts of the source may be magnified more than other parts. 
In the case of stars, for
example, the limb can be brightened relative to an unmagnified source.
Since the light from the limb comes from different temperature profiles in 
the star compared 
to the center, this will affect the spectrum \citep[e.g.][]
{valls-gabaud:95,loeb:95} For example, a spectrum of the limb of the Sun 
has Balmer lines with weaker wings, while lines of other elements
can be either weakened or strengthened relative to the center \citep{hastings:73,hale:07}.

Recently, observers using large telescopes have published abundance
ratios in dwarfs in the Galactic bulge from spectra obtained while the
sources were highly magnified.  A number of these have turned
out to have super-solar metallicities, including OGLE-2006-BLG-265S
\citep{johnson:07}, MOA-2006-0BLG-099S \citep{johnson:08},
OGLE-2007-BLG-349S \citep{cohen:08}, MOA-2008-BLG-310S and
MOA-2008-BLG-311S \citep{cohen:09} and OGLE-2007-BLG-514S \citep{epstein:09}. Some are metal-poor including the subgiant
OGLE-2008-BLG-209S \citep{bensby:09a} and the dwarf OGLE-2009-BLG-076S
\citep{bensby:09b}, which, at [Fe/H]$=-0.76$ is the most metal-poor
dwarf/subgiant observed when microlensed. For a sample of eight
microlensed dwarfs, \citet{epstein:09} found that a K-S test
gave a 1.6\% chance that the dwarfs were drawn from the same
metallicity distribution function (MDF) as the giant MDF from \citet{zoccali:08}. \citet{bensby:09c} report
results for a total of 13 dwarfs, including new results for 
more metal-poor dwarfs, so
clearly the comparison between the giants and the dwarfs is an
evolving topic. \citet{cohen:08} proposed that mass-loss on the 
giant branch prevents
some more metal-rich stars from becoming red giants, similar to the
mechanism suggested by \citet{kalirai:07} to explain the low-mass He
white dwarfs in the metal-rich cluster NGC 6791. However, this
explanation predicts that the red clump MDF is biased to more
metal-poor stars because of metal-rich stars skipping the horizontal
branch phase and that the red giant luminosity function should drop
more than expected from theoretical models, predictions that do not
agree with the giant data \citep{zoccali:08}. Another explanation is
that there are systematic differences between giant and dwarf
abundance analyses.  \citet{cohen:09} calculated that a systematic
offset of 0.10 dex between the dwarf and giant metallicity scales
would give the mean metallicity offset a significance of 2-$\sigma$
and larger systematic offsets would obviously decrease the
significance further still. Finally, \citet{zoccali:08} suggested that
the spectra of microlensed dwarfs could be affected by differential
magnification sufficiently that the usual analysis of the dwarfs,
which does not take this into account, could lead to biased answers and help
explain the possible discrepancy.  This last suggestion can be tested
by comparing the answers obtained from synthetic spectra with and
without differential limb magnification.

While most microlensing events follow the lightcurve of a point source, about
$\sim$3\% \citep{witt:95} of events show finite source
effects. This fraction is even higher for high-magnification events
that are targets of the current generation of dwarf studies. Of the
eight published events for which dwarf spectra were obtained, 
we know that at least three of the events were affected by finite source
effects, namely OGLE-2007-BLG-514, OGLE-2007-BLG-349, and MOA-2008-BLG-310.  
We need to determine the size of the
effect that differential limb magnification (DLM) 
 has on the spectra and the measurement of the
effective temperature (\teff), gravity (\logg), metallicity ([Fe/H]),
microturbulent velocity ($\xi$) and abundance ratios to correctly
interpret these events.

In addition to probing the chemical evolution of the bulge,
the accuracy of the measured \teff{} from the spectrum is important for 
testing the method by which colors of source stars in microlensing events
are determined. Using a metallicity and a \teff{} derived from 
the spectrum, we can predict the color of the star using relations
between color and \teff, such as that by \citet{ramirez:05}, and compare
with the color estimated using standard microlensing techniques, which
rely on the offset of the star from the red clump.
The results so far indicate that if the color of the 
red clump is $(V-I)_0$=1.05, the \teff s derived spectra are in agreement
with \teff s from colors. 

Much work has been done on the effects of DLM of
the disk of giants during a microlensing event because these
events are easier to find and are longer-lasting than similar events
in dwarfs. \citet{valls-gabaud:98}
predicted the effects on the spectrum of a giant, in particular
the H$\beta$ and CO lines, by fitting simple analytic expressions
to the limb profiles for both continuum and lines 
from the \citet{kurucz:92} models. 
He found that the equivalent widths could change by
20\% over the course of the event. \citet{heyrovsky:00} computed more
realistic models avoiding the use of linear approximations. 
They calculated contribution functions for
lines, which indicate where the lines are formed, and illustrated
the range of behavior individual spectral features will have 
because of varying center-to-limb profiles. 
\citet{gaudi:99} pointed out that binary lens events had particular 
advantages for resolving stellar surfaces, for example the
fact that the second caustic
crossing can be predicted if the event is monitored, and therefore
carefully observed.

Thanks to intensive monitoring by observers, the effect of the size of
the source on microlensing events has been observed many times.
Finite source effects were observed for the giant MACHO 95-30 \citep{alcock:97},
including variations in the equivalent widths of H$\alpha$ and the TiO
bands. The bulge K3 giant star EROS
BLG-2000-5S was highly magnified by a binary lens. Because the
caustics of a binary lens are symmetric, the timing of the second
caustic crossing was predicted to high accuracy and allowed intensive
follow-up during this event. Both \citet{albrow:01} and
\citet{castro:01} found decreases in the strength of the H$\alpha$
line when the limb was more highly magnified relative to the
center. \citet{afonso:01} monitored the same event photometrically,
measured the color of the limb, and found that the outer 4\% of the
limb showed strong H$\alpha$ emission, which they ascribed to the
presence of a chromosphere.  
\citet{albrow:99} performed the first limb-darkening
measurement of a bulge giant for the event MACHO 1997-BLG-28, using
photometry in the $V$ and $I$ bands.  Other limb-darkening measurements 
were done for giants including MACHO 1997-BLG-41 \citep{albrow:00}, and 
OGLE-2002-BLG-069S \citep{kubas:05}, which the authors found
agreed with
a linear limb-darkening law. An opposite result was found by 
\citet{cassan:06} who obtained photometry for the K giant
OGLE-2004-BLG-254S and derived limb darkening coefficients in the $I$
and $R$ bands. Combining these results with other microlensed-based
limb darkening measurements of K giants and comparing them to the
predictions of ATLAS9 models, they found a disagreement, which they
suggested was the result of the inadequacy of the linear limb darkening
law. Time-resolved spectra of
OGLE-2002-BLG-069S, a G5III star, by \citet{cassan:04}, showed the
power of this technique for probing the stellar atmosphere. In this
case, the changes during the event 
in  wings of H$\alpha$ line, which are formed deep in the 
atmosphere, agreed with model predictions while the core of the
line showed an emission peak, which again can only be explained by a 
chromosphere. That the
temperature structure in the outer layers is not well understood for
this event was also found by the analysis of
\citet{thurl:06}. These studies show the 
potential to investigate the atmospheric structure of giants.

Studies of dwarf stars have been much rarer.
\citet{afonso:00} and \citet{abe:03} reported the only measurements 
of limb darkening in dwarfs, for MACHO 98 SMC-1, a main-sequence A star
and MOA 2002-BLG-33S, a solar-type star, respectively.  
The most relevant work to this paper is the analysis of \citet{lennon:96}
of the event MACHO 96-BLG-3. They took three exposures when the source,
a dwarf star, was moving across a caustic. These R$\sim$1100 spectra 
with signal-to-noise (S/N) of 25-100 did not show any
convincing cases of profile variability. 
\citet{lennon:96} compared
the observations of this star with both a library of high-resolution,
high S/N spectra of F and G stars observed at Calar Alto and 
a grid of synthetic spectra, and derived stellar parameters of
\teff=6100K from the H$\alpha$ line, \logg=4.25 from the \ion{Mg}{1} triplet
and a metallicity ([M/H]) between 0.3 and 0.6 from fits to regions 
with many \ion{Fe}{1} and \ion{Ca}{1} lines. They calculated the 
expected deviations from the unmagnified spectrum for this event, and
found that the expected change in the line profiles for the
three spectra was $\leq$ 1\%, while the expected change in the
continuum was $\leq$ 2\%. Given the small changes expected and
the S/N of the spectra, it is not surprising that no changes were observed,
and that the derived atmospheric parameters of the star would also not be
affected. Indeed, it is not surprising that most changes are small.
The amount of magnification depends on the distance from the
lens, but the emitted spectrum is the same for an entire annulus (for
a spherical star). Because the distance from the lens varies around the
annulus, the average magnification of a spectrum at a particular
annulus is smaller than the largest magnification of an particular spot would suggest. In
addition, while the spectrum of the star increasingly changes from
center to limb, the intensity of the limb is lower than the
center, by factors of a few. Therefore
it is difficult to overcome the influence of the light
coming from the central regions of the star in the disk-averaged light.

The work discussed above focused on learning about the
atmosphere of the star and showed that finite source
effects cause changes in the spectrum of the star, which
can be inverted to give temperature as a function of depth. 
Because the same effects that make these extreme events interesting
as probes of the atmosphere will also change the spectrum that is
analyzed for elemental abundances, the goal of this paper is
to determine the effect of DLM on the
\teff, $\xi$, \logg, [Fe/H] and abundance ratios using the standard
spectroscopic techniques, such as equivalent width analysis. 
Therefore, we will be concerned with mimicking standard abundance
analysis as closely as possible, rather than exploring the full
range of knowledge that can be gleaned from time-resolved spectroscopy
of microlensed events. 

\section{Magnification Profiles during Microlensing Events}

\begin{figure}
\includegraphics[width=8cm]{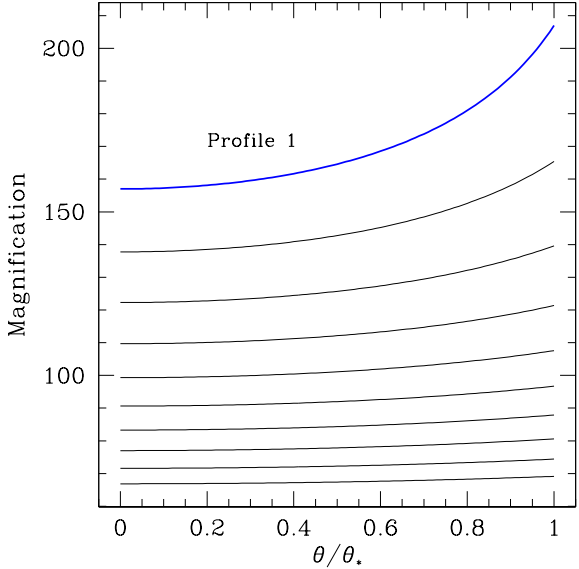}
\caption{Differential limb magnification profiles for Event
A. The bold line shows Profile 1, which represent the most extreme differential
magnification for this event, and is the profile from the first
15 minutes of Event A. At most, the limb is magnified by $\sim 30\%$ more
than the center.}
\label{fig:moa310}
\end{figure}

Our work was motivated by two events, MOA-2008-BLG-310
and OGLE-2007-BLG-514, 
which are illustrations of the kind of effects seen.
For MOA-2008-BLG-310, the lens did in fact transit the source,
and there were small perturbations during this
transit caused by a planetary companion in the lens \citep{janczak:09}.
However, the spectrum analyzed by \citet{cohen:09} began
at UT 22:51, which was 21 minutes after the end of the transit,
when the lens and source center were separated by $z=1.22$ source
radii. Figure~\ref{fig:moa310} shows the annulus-averaged profiles for a point lens
at 15 minute intervals for the MOA-2008-BLG-310 geometry, beginning
at $z=1.2$ and continuing as the lens and source moved
further apart. The magnification is given as a function of $\theta$, which
is the angle between the normal to the
stellar surface and the line of sight to the observer. To determine
This look at the later stages of MOA-2008-BLG-310
will be referred to as Event A. The start time for the first profile 
is three minutes before the \citet{cohen:09} observations
began, and thus serve as a direct measurement of the size of the
effect of DLM on those abundances, but they also serve a broader
purpose.

In this particular case, we know that $z=1.22$ at the
time of the observations because the source
size was earlier detected by observations (from Africa) of
a direct source crossing. However, if the closest separation between
source and lens during the entire event (i.e., the impact parameter) had
been $z_0=1.22$, there would have been no source crossing at any time.
If the impact parameter is sufficiently large, then it becomes
impossible to determine the source radius from the lightcurve. In
such a case, only upper limits can be placed on the
magnitude of DLM and profiles of the DLM as a function of $\theta$ cannot
be derived.

However, for $z_0\leq 1.2$, the effects of the finite-source
size on the light curve are sufficiently pronounced to measure
$z_0$. Thus, the top curve in Figure~\ref{fig:moa310} represents the most
extreme case of DLM that would occur without being noticed, and
thus serves as a {\it general} check on ignoring DLM when
it is not detectable from the light curve.

\begin{figure}
\includegraphics[width=8cm]{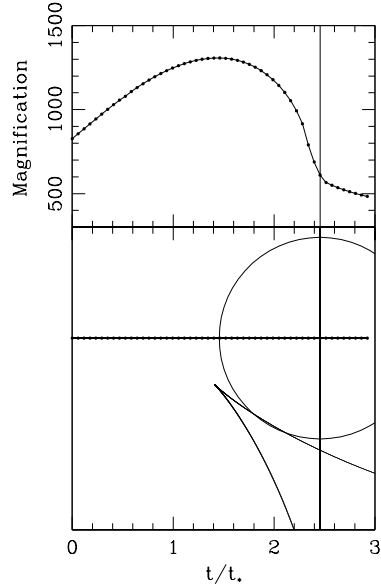}
\caption{The top panel shows, for Event B, the source-averaged magnification
as a function of time, measured in units of the source-radius
crossing time.  The bottom panel shows the geometry of the event
at a moment when the source (circle) is just exiting the caustic
(acute-angled structure), which is a contour of formally infinite
magnification.  The source-center position is marked by small
circles at 5 minute intervals. As the source passes over the
caustic, it is differentially magnified, with first the limb,
then the center, and finally the limb being the most magnified.}
\label{fig:traj}
\end{figure}

\begin{figure}
\includegraphics[width=8cm]{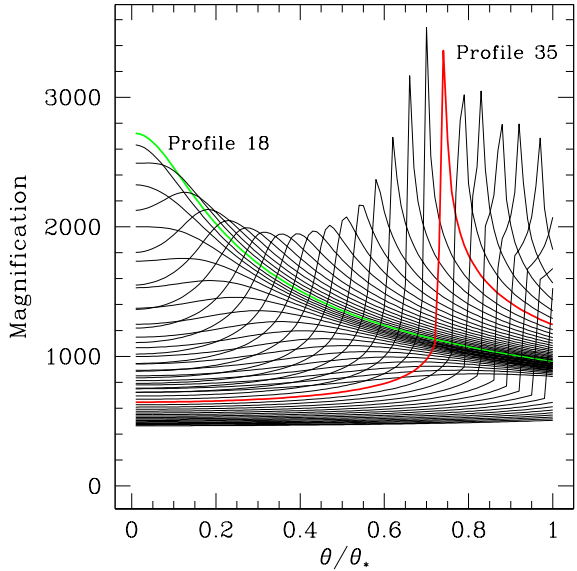}
\caption{The magnification of the disk of a dwarf as a function of
radius at each interval marked in Figure~\ref{fig:traj}. 
We see in the most extreme
cases that parts of the disk are magnified up to 5$\times$ more than
other parts. The two profiles that are examined in more detail
in the text are for time step 18 (bold line peaking at
$\theta=0$)  and time step 35 (bold line peaking at $\theta=0.75$)}
\label{fig:ogle514}
\end{figure}

The other case represents a more extreme event,
inspired by OGLE-2007-BLG-514, and will be called Event B. 
Here, the source trajectory crossed a cusp from a binary lens 
and produced extreme magnification variations (Figure~\ref{fig:traj}). Once the
parameters are selected, it is straightforward to compute the
magnification at each time and for every point on the source plane. We
calculated the mean magnification in concentric rings with annuli equal
to 0.01 source radii using the ``loop-linking'' technique
\citep{dong:06}. The profiles are spaced at five minute intervals for
this event. Actual observations are always longer, usually 5-6$\times$ longer
in order to get enough S/N in each exposure to reliably extract the
spectrum. Spectra at 5 minute intervals could only be obtained for
dwarfs magnified to apparent magnitudes that have not been seen in an
event to date, so in reality microlensing spectra will smear out these
profiles and dilute their effects. 
Figure~\ref{fig:ogle514} shows 51 different profiles that
occurred during this event.  We note that in this case we have ample
notice from the lightcurve that finite source effects are important
and we could use that information to derive the magnification profile
at the time spectra were taken and make model spectra that include the
effects of DLM.  However, in this paper, we consider the cases that
ignored  finite source effects on the spectrum to measure the size
of errors that this induces on the parameters, metallicities and abundance
ratios.

\section{Synthesized Spectrum}

We explored the changes to the spectra of dwarfs 
caused by the magnification profiles given in \S 2. We
interpolated a set of model atmospheres from the Kurucz-Castelli grid with new
opacity distribution functions (\citet{castelli:03} 
\footnote{Available at http://wwwuser.oat.ts.astro.it/castelli/grids.html}. 
The models are solar-metallicity and are spaced every 100
K from 5000K to 6200K. This temperature range covers the range of
temperatures for dwarfs observed in the bulge. Cooler dwarfs ($<5000$
K) are unlikely to ever be targeted because of their intrinsic
faintness and because if the dwarfs are as cool as the giants, many of
the advantages of measuring abundances in dwarfs rather than giants
(such as reduced blending) no longer apply. Although the majority of
the bulge population is old \citep{ortolani:95}, we wanted to explore
a large
temperature range of main-sequence/main-sequence turnoff stars, and 
therefore we adopted \logg{} values from a Yale-Yonsei isochrone
\citep{yy} of 4 Gyr. At this age, stars with \teff =6000K have two
possible values for \logg, because they are at the turnoff, while
dwarfs with 6100K and 6200K are still present because of the blue hook
in the isochrone. The two 6000K dwarfs allow us to measure the
effect that small changes in \logg have on the resulting spectra. Table
1 lists the temperatures and \logg's for the dwarfs.

\begin{deluxetable}{lrrr}
\tablenum{1}\label{Tab:models}
\tablewidth{0pt}
\tablecaption{Parameters for Model Atmospheres}
\tablehead{
\colhead{\teff{}(K)} & \colhead{\logg} & \colhead{[Fe/H]} & \colhead{$\xi$ (km/s)}} 
\startdata
 5000 & 4.60 & 0.00 & 1.500\\
 5100 & 4.59 & 0.00 & 1.500\\
 5200 & 4.57 & 0.00 & 1.500\\
 5300 & 4.55 & 0.00 & 1.500\\
 5400 & 4.53 & 0.00 & 1.500\\
 5500 & 4.51 & 0.00 & 1.500\\
 5600 & 4.48 & 0.00 & 1.500\\
 5700 & 4.46 & 0.00 & 1.500\\
 5800 & 4.43 & 0.00 & 1.500\\
 5900 & 4.39 & 0.00 & 1.500\\
 6000 & 4.33 & 0.00 & 1.500\\
 6000 & 4.06 & 0.00 & 1.500\\
 6100 & 4.01 & 0.00 & 1.500\\
 6200 & 3.98 & 0.00 & 1.500\\
\enddata
\end{deluxetable}

\subsection{Method of Synthesizing Spectra}

We focused our attention on three 200 \AA{} sections of the spectrum,
centered on H$\alpha$ (6460 \AA -- 6660\AA), H$\beta$(4757\AA --
4957\AA) and the \ion{Mg}{1} triplet (5067 \AA -- 5267 \AA). These
regions have both strong lines, which are expected to show the largest
variations, as well as a number of weaker lines whose equivalent widths
would be used in an abundance analysis. The wavelength range from blue
to red also ensures that detectable unblended lines can be found for
both cooler stars, which have crowded blue regions, and hotter stars,
which tend to have weak lines in the red. The H$\alpha$ and
H$\beta$ lines are also used as a temperature indicator in hotter
dwarfs. For each model atmosphere, we used
Turbospectrum \citep{turbo} to generate intensities I($\theta$) 
at 100 values of
$\theta$. To determine
the total flux from the star, we added up the intensities coming from
annuli from the center to the radius R of the star

\begin{equation}
F=2\pi\int_0^R I(\theta) r dr
\end{equation}

\noindent{\citep[see e.g.,][]{mihalas:78} Using $r=R sin\theta$, for the unmagnified case, the observed
flux can be obtained by numerical integration of the
equation}

\begin{equation}
F=2\pi\int_0^\frac{\pi}{2} I(\theta) cos\theta sin\theta {\mathrm d}\theta
\end{equation}

This gives the same answer as when Turbospectrum outputs a flux, 
rather than intensities.
For the magnified cases, each annulus was multiplied by
its magnification factor before integration.
The spectra were smoothed to a FWHM of 0.11\AA, or R$\sim$ 45,000,
similar to the resolution at which the dwarfs are observed.

\subsection{Comparison of Spectra}

To investigate the size of possible systematic effects on the
abundances in dwarf stars, 
we wished to determine the maximum effect on the spectroscopic analysis,
and therefore begin by identifying the cases for which the spectra deviate
the most from the unmagnified case.
We took the ratio between the magnified and the unmagnified
spectra, renormalized the spectra and then calculated the
rms. 

For Event A, neither the profiles nor the rms varies much, but the
largest deviations are found for the profile calculated for
the first time step, indicated
by the bold line in Figure~\ref{fig:moa310}. 
For Event B, the rms is highest, as expected, 
when the limb is magnified by a high factor or when
the center region is magnified the most (bold lines in 
Figure~\ref{fig:ogle514}). These occur for the 18th and 35th profiles
calculated, corresponding to 85 and 170 minutes after Event B began. 
We will use these three
profiles as examples to calculate the size of the effects on the
abundances. The indicated case from Event A will be called Profile 1 and
the two indicated cases from Event B will be called
Profile 18 and 35.

\begin{figure*}
\includegraphics[width=12cm,angle=270]{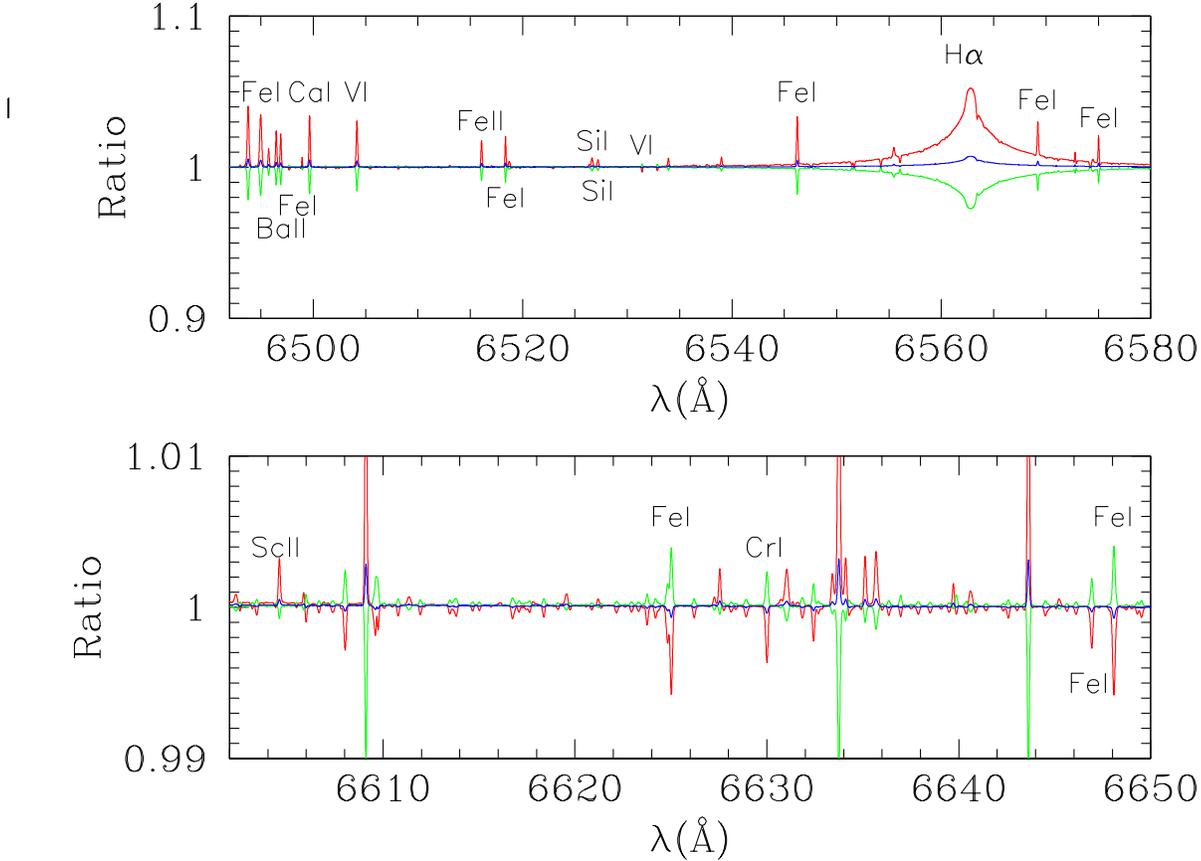}
\caption{Ratio of the spectra in the magnified cases to those in the
unmagnified case for the 5500K model. 
The red line is for Profile 35, the green line for
Profile 18 and the blue line is for Profile 1.  Top: part of the H$\alpha$
spectrum, showing the stronger lines tend to be weaker for Profile
35 than for Profile 18, and in both cases the effects are much
stronger than for Profile 1.
 Bottom: part of the H$\alpha$ spectrum, focusing on weaker
lines, which tend to be weaker for Profile 18 than for Profile 35.
Again, Profile 1 shows the smallest effects.}
\label{fig:compspec}
\end{figure*}

Figure~\ref{fig:compspec} illustrates the changes to a spectrum by showing the
ratio of the unmagnified spectrum to the spectra created using the three
different magnification profiles for the 5500K dwarf. 
The changes are small, but vary from line-to-line based
on the location of the line-forming region. In general, 
the strong lines, including
H$\alpha$ and H$\beta$, become weaker for Profiles 35 and 1,  
for which the limb is magnified relative to the center, while the reverse
is true for the spectrum from Profile 18. The opposite behavior is
seen for most of the weak lines. As expected, the deviations
are largest for the two cases where the magnification profile is more
extreme and much smaller for Event A, where
the differential magnification effects were $\sim 30\%$, rather
than $\sim 500\%$.

\section{Abundance Analysis}

The goal of this paper is to determine the practical effect of finite
source effects on abundance measurements in microlensed dwarfs. To do
this, we followed the usual steps used to analyze these
spectra. Because differential reddening across the face of the bulge
leads to uncertainty in the colors and magnitudes of stars in the
bulge, analyses of microlensed dwarfs have relied on spectroscopic
methods of deriving atmospheric parameters, rather than color or
apparent magnitude-based methods. DLM can impact the derived
abundances in the source star by changing the EWs of the lines and by
changing the atmospheric parameters derived from those EWs. To examine
these effects and compute the total effect of DLM on the abundances,
we first measured EWs in the synthetic spectra for the unmagnified
case and Profiles 18, 35 and 1. We ran all sets of EWs
through the original model atmosphere to see what changes. We
found small changes in the diagnostics used to determine \teff, \logg,
and $\xi$, in addition to changes in the abundances. We first discuss
the magnitude of the changes demanded by the magnified spectra on the
atmospheric parameters, considering each one in isolation. While the
exact magnitude of the changes depends in detail on which lines are
used for any particular study, the results here will give a general
indication. Looking at the changes in each parameter separately shows
the size of the effect if only some of the parameters are determined
from the spectra. However, because deriving the atmospheric
parameters usually depends on the other
parameters, we next consider the total effect of the accumulated
atmospheric parameter changes plus EW changes. Once again, depending
on the line list used, different correlations between the parameters
may be found, but the overall tendencies can be generalized.

\subsection{Equivalent Width Measurement}

In measuring EWs on the model spectra, we encountered the same
concerns about continuum placement and blending as in measuring EWs on
observed spectra. Because of our desire to determine differences
caused by differential limb magnification, we focus on measuring EWs
in a consistent manner. We first used IRAF\footnote{IRAF is
distributed by the National Optical Astronomy Observatories, which are
operated by the Association of Universities for Research in Astronomy,
Inc., under cooperative agreement with the National Science
Foundation.} to do the continuum division, keeping the parameters of
the fit, such as the order of the polynomial and the high and
low-reject sigmas, the same for both magnified and unmagnified
spectra.  Then we measured the EWs without adjusting the continuum
interactively, using SPECTRE (C. Sneden, 2007, private communication)
to fit Gaussian profiles to the spectra.  Our initial linelist was
selected using the solar atlas of \citet{charlotte} as a guide to
unblended lines, but this was not completely foolproof at the lower
resolution (compared to the solar atlas) of the synthesized spectra,
and over the large temperature range spanned by the models. Therefore
we checked the list by determining the abundances from the lines for
the unmagnified spectra. Any line that deviated by more than 0.1 dex
from the mean is eliminated. Also eliminated are lines with EW $>
150$m\AA{} in the unmagnified case, because these lines are usually
eliminated in high-resolution analysis due to their sensitivity to
damping parameters, continuum placement, blending in the wings and
incorrect temperature structure in the outermost layers of the model
atmospheres. For each temperature, the same lines were measured for
each magnification profile; however, because the strength of lines
varies with temperature, the linelist changes for different
temperatures.  Because the same transition information, such as
$gf$-value, was used both to synthesize the spectra and to calculate
the abundance, there are no uncertainties introduced in this process
by atomic data uncertainties.

\begin{figure}
\includegraphics[width=8cm]{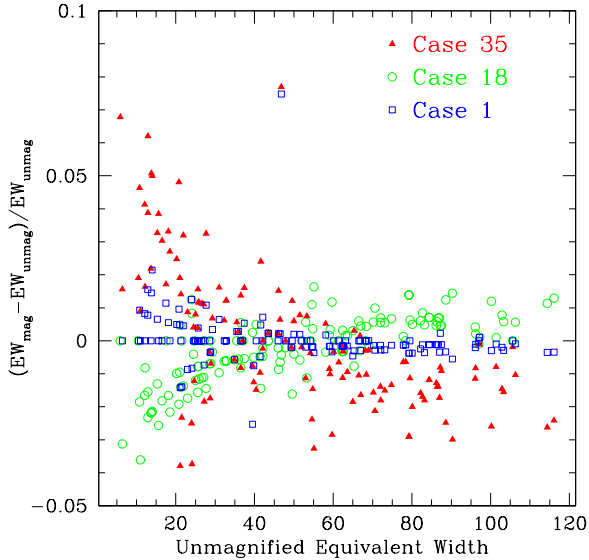}
\caption{A comparison of the EWs measured in the magnified spectra to
those in the unmagnified spectrum for the 5500K model. }
\label{fig:ew5000}
\end{figure}

\begin{figure}
\includegraphics[width=8cm]{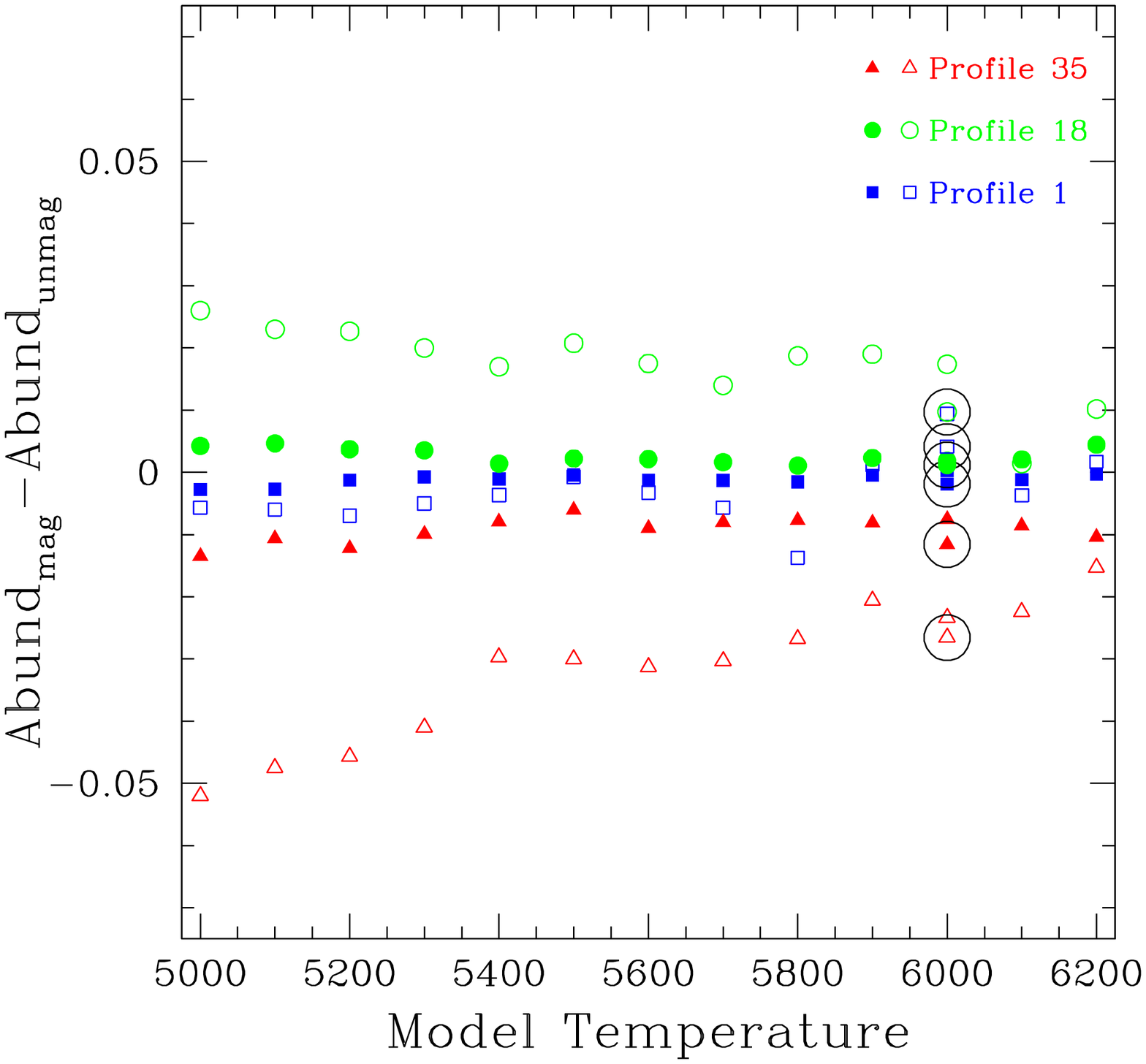}
\caption{The change in \ion{Fe}{1} abundances (filled symbols)
and \ion{Fe}{2} abundances (open symbols) that result from changes
in the EWs alone. Different linelists and continuum
division for different temperature models lead to more bumpiness
in the relation with temperature than otherwise. Large circles
show the lower gravity
6000K model points.
Overall, it is clear that the changes in the EWs in themselves do not
result in large changes to the abundances. For Profile 18, for which 
the center was magnified more than the limb, analyzing the EWs assuming
the unmagnified model results in a increase of the derived metallicity,
while the reverse is true for Profile 35. Very small changes are
seen for Profile 1, as expected given the small deviations seen 
in Figure~\ref{fig:compspec}}
\label{fig:ewabund}
\end{figure}

In Figure~\ref{fig:ew5000}, we compare the difference in EWs for the three 
magnification profiles under study. We find that, as expected, the
smaller differences between limb and center for Profile 1 result
in smaller changes in the EWs. As seen in
Figure 4, weaker lines are mostly strengthened and strong
lines mostly weaken for Profile 35, while the opposite is true for Profile 18.
This emphasizes the point that the magnitude of the changes for
a particular analysis will depend on the distribution of EWs that
are being used. We then ran the EWs through Turbospectrum, using the
original model atmosphere. This comparison isolates the
effect of the EW changes on the abundances, before we consider
changes caused by new atmospheric parameters, and would reflect the
total changes if the analysis relied on non-spectroscopic methods for
atmospheric parameter determination,
such as color of the unmagnified star for \teff{} and 
position on the CMD for gravity. The changes in
abundances are shown in Figure~\ref{fig:ewabund}, and show
the magnitude of $\delta$ log$\epsilon$ is generally < 0.05 dex.

\subsection{\teff}

There are two standard ways of determining the temperature of a
star from its spectrum, without relying on its color. The first
uses the wings of the Balmer lines, while the second requires
excitation equilibrium for \ion{Fe}{1}, so that there is no
trend in a plot of lower excitation potential to abundance derived
for each line. 

\begin{figure}
\includegraphics[width=8cm]{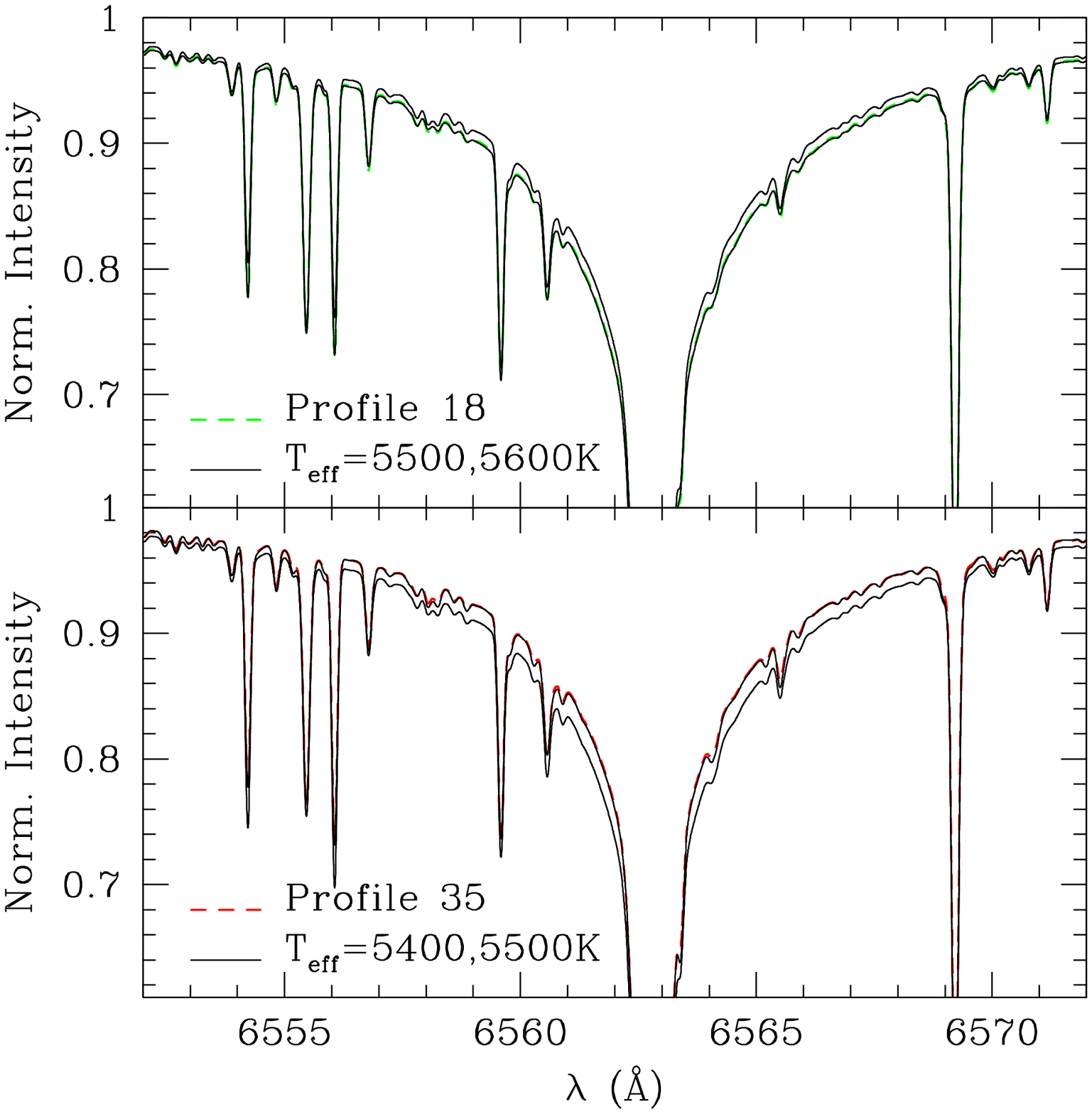}
\caption{Comparison of unmagnified spectrum at H$\alpha$ with magnified
spectra for Profile 18 (top) and Profile 35 (bottom). The magnified
cases are shown for \teff=5500K. In the top panel, the unmagnified
cases shown are for \teff=5500K and \teff=5600K. The slightly broader
wings from the Profile 18 spectrum show that we would calculate
a higher temperature (~100K) from this spectrum if we ignored DLM. This
agrees with the larger contribution from the center of the star, because
its spectrum features stronger Balmer lines.
In the bottom panel, the unmagnified cases shown are for
\teff=5500K and \teff=5400K. Here we would have calculated
a lower temperature, which is in agreement with the
prominence of the limb, with its weaker Balmer lines, in this
particular profile.  }
\label{fig:halpha}
\end{figure}

H$\alpha$ is well-known to show the effects of DLM. Figure~\ref{fig:halpha}
compares the H$\alpha$ profiles of the unmagnified case to 
the most extreme magnified cases (Profile 18 and 35) 
for the \teff=5500K model. It is clear that
the H$\alpha$ profiles are different in the magnified models, and
a different temperature would be derived if DLM were not
taken into account. By comparing the magnified cases with the unmagnified
profiles of the different temperature models, we determine that
the \teff{} we would derive would be 75K or 100K hotter
for Profile 18 and would be 75K or 100K cooler for Profile 35. This
is consistent with the magnification profiles for these two cases,
because Profile 18 has the center, where we see to hotter temperatures,
magnified more than the limb, while the opposite is true for Profile 35.

\begin{figure}
\includegraphics[width=8cm]{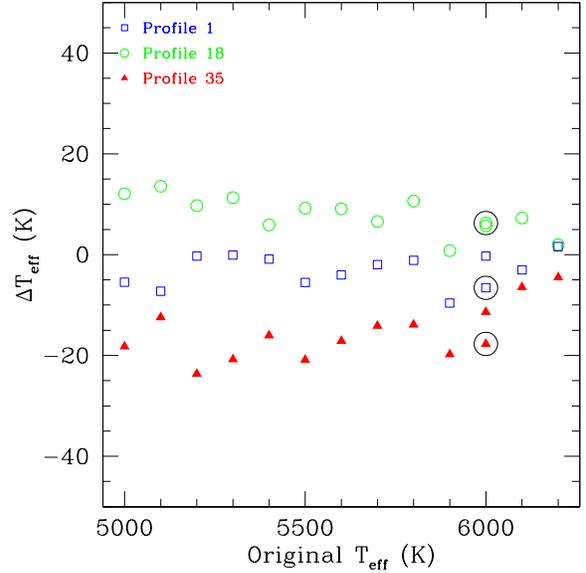}
\caption{The change in \teff{} required to obtain the same excitation
equilibrium as in the unmagnified case. Profile 18 
requires higher temperatures, because the center is magnified
relative to the limb, while Profiles 35 and 1
require lower temperatures. Overall, the temperature changes are
larger at lower temperatures. Large circles mark the lower gravity
6000K model. The original gravity of the star makes only
a very small difference. }
\label{fig:newteff}
\end{figure}

Determining the temperature via the wings of the H$\alpha$ line requires
high S/N data and becomes increasingly difficult at lower temperatures.
Therefore, the more common method, used by all the papers on microlensed
dwarfs discussed in the Introduction, is excitation
equilibrium. We found that the slope of the line, determined
by a least-squares fit, in the excitation 
potential-log$\epsilon$(Fe) plane changes if we used the
EWs measured from the DLM spectra. Because of errors
in continuum division and blending, the slope of the line in the
unmagnified case, which theoretically should be zero, deviates slightly
from that value. Therefore, to calculate the change in \teff{} from this method
caused by the effects of DLM on the spectrum, we calculated the change in
temperature needed to get the slope back to that measured in the
unmagnified case. Because we found the temperature shifts are small, 
we ran a series of model atmospheres with the
temperature decreased by 25K, found the new slope using the
magnified EWs, and used the results to calculate the necessary temperature
change. The results are illustrated in Figure~\ref{fig:newteff}. 
We also did the test with +25 K and got the
same answer to within 1 degree. 

\subsection{Microturbulent Velocity}

\begin{figure}
\includegraphics[width=8cm]{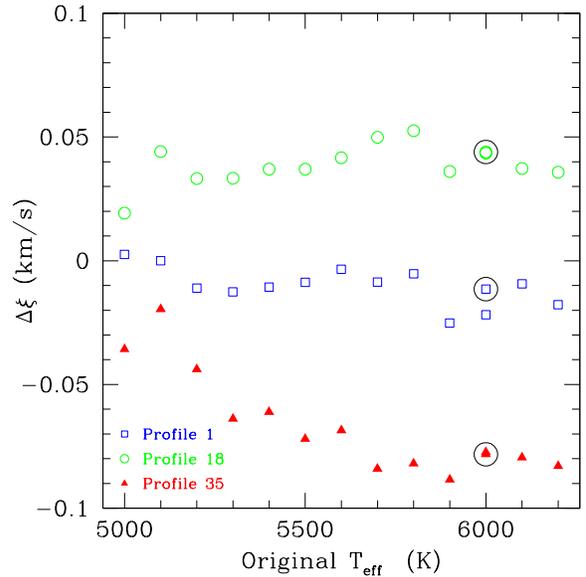}
\caption{The changes in $\xi$ necessary to
have the slope in the RW/\ion{Fe}{1} plane be the same
as in the unmagnified case for which only $\xi$ is allowed to change.
Large circles mark the lower gravity 6000K model.}
\label{fig:newxi}
\end{figure}

The microturbulent velocity 
is determined for 1-D analysis by ensuring that the slope for
the derived abundance as a function of reduced EW for lines of an
element is zero. Usually, the large number of \ion{Fe}{1} lines make
it the element of choice for this test, and this was the case for the
analyses of the microlensed dwarfs. We compared the slopes of the
lines in the \ion{Fe}{1} vs. reduced EW (RW) 
plot in the magnified case to that in
the unmagnified cases using the Fe abundances determined with 
the original model atmosphere in both cases. There is a change in the
slope of the line in the magnified cases, because their EWs are
affected by DLM.. As with the \teff , the slope of the line in the
unmagnified case is not exactly zero. We run our EWs and original
model atmospheres through Turbospectrum, this time with $\xi$ set to
1.4 km/s and use the change in the slope for 0.1 km/s to interpolate
the change in $\xi$. The magnitude of the change is always $< 0.1$ km/s
(Figure~\ref{fig:newxi}).

\subsection{\logg}

\begin{figure}
\includegraphics[width=8cm]{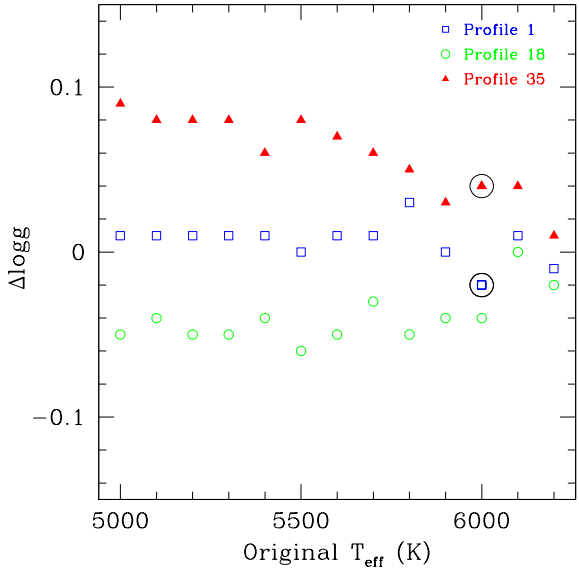}
\caption{The changes in \logg{} for each model necessary to bring
\ion{Fe}{1} and \ion{Fe}{2} back into agreement, keeping the temperature
at the original temperature and $\xi$ at the original $\xi$.
Large circles mark the lower gravity 6000K model.}
\label{fig:newg}
\end{figure}

For the analysis of the microlensed dwarfs, gravity is 
determined spectroscopically by demanding 
that \ion{Fe}{1} and \ion{Fe}{2} (or other ions) give the same abundance. Obviously, changes
in the EWs of these lines as well as changes in \teff{} and $\xi$ will
result in a change in the derived \logg. In Figure~\ref{fig:newg}, 
we show the changes in \logg{} that are necessary to get agreement
between \ion{Fe}{1} and \ion{Fe}{2} considering the changes in
EW alone. 

\subsection{Abundances}

In the previous subsections, we have considered the changes
necessary in \teff, $\xi$ and \logg{} if we kept the other parameters
equal to those of the original model and only modified the one under consideration.
This is appropriate when one model atmosphere parameter is chosen
independently of the others.
However, when deriving a model atmosphere, changes in one parameter
will usually propagate and lead to total changes that can be
larger or smaller than the partial derivatives suggest. In this section,
we now consider the total effect on the calculated abundances, 
both from all the correlated changes in  determining 
of the model atmosphere parameters and also from the changes in the EWs of the
elements being measured. 

For our particular linelist, \teff, \logg{} and $\xi$ all affect
the slope of the excitation potential vs. \ion{Fe}{1} abundance plot,
the slope of the reduced equivalent width vs. \ion{Fe}{1} abundance plot
and the difference in abundance between \ion{Fe}{1} and \ion{Fe}{2}
lines. Therefore, using the partial derivatives we calculated above,
we simultaneously solve for the changes in \teff, \logg{} and $\xi$ that
resulted in the slopes of the two lines and the difference in
the two ions being the same as in the unmagnified case. The changes
in the atmosphere models from the original parameters in Table 1 are
listed in Table 2. The changes are given as magnified parameters
$-$ unmagnified parameters.
Finally, we interpolated a model with those parameters and run the
EWs through that model to calculate the final change in the abundances.
We checked that the final model has the excitation, equivalent
width and ionization equilibrium to within the scatter (e.g., < 0.01 dex
difference in ionization equilibrium) expected considering the
effects of rounding off temperatures, gravities and velocities. As these
are interpolated models to begin with, these changes will never
be that precisely determined. 
We note that the metallicity of the model atmosphere will have
an effect on the abundances. However, the changes in log$\epsilon$(Fe)
based on \ion{Fe}{1} lines, which is how the metallicity of the atmosphere
is usually set, are small enough that we will ignore this last effect
for the purposes of the paper. For example, the change of 0.02 dex in
\ion{Fe}{1} for the 5600K model with Profile 35 would lead to a
subsequent change for 0.006 dex in \ion{Fe}{1} if folded back into
a new model. Of course, the detailed analysis of a particular event
could take these into account.

\begin{figure*}
\includegraphics[width=16cm]{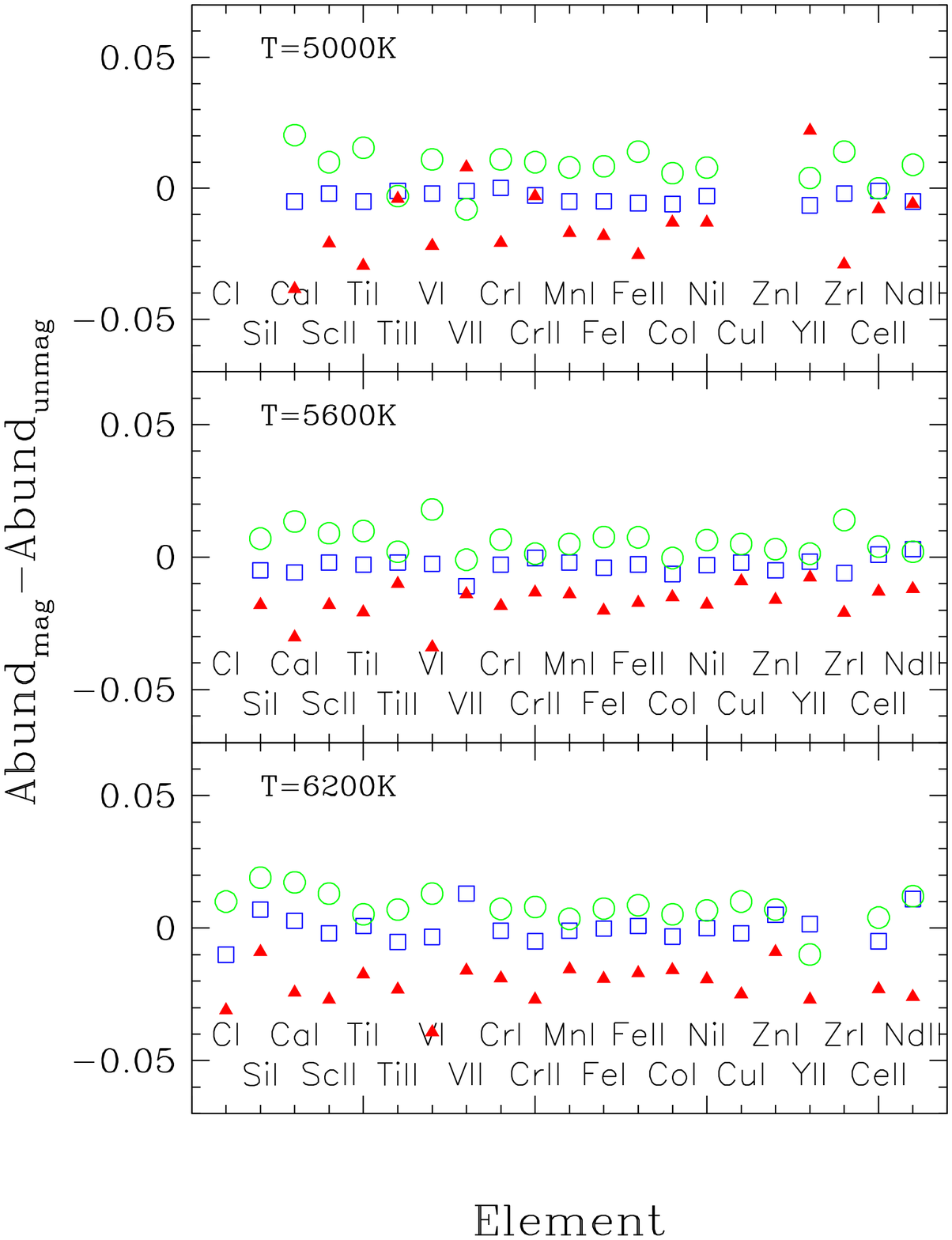}
\caption{The change in abundance (unmagnified-magnified) 
from the unmagnified case for
three profiles studied in detail in this paper. {\it Top:} 5000K 
model {\it Middle:} 5600K case {\it Bottom:} 6200K case. Not all
elements were measured for each temperature. We see that when
DLM is not taken into account when analyzing the spectrum, the
abundances can be in error by up to 0.05 dex. }
\label{fig:newabund}
\end{figure*}

Figure~\ref{fig:newabund} shows the resulting changes in the
log$\epsilon$ for the elements considered here for 5000K, 5600K and
6200K. This figure illustrates the small magnitude of the changes
expected from DLM in the dwarfs. Also, the changes in abundances
are not always in the same sense. Depending on what part of the
source is most highly magnified, the change caused by not
taking DLM into account can either be
an increase or decrease over the true abundances in the star.
For some ions, such as \ion{Fe}{2}, the changes using the
new parameters are smaller than using the new EW with the unmodified
atmosphere (seen by comparing Figure 6 and Figure 11). For
others, such as \ion{Fe}{1}, the changes with the new parameters
are larger than seen just using the new EWs.

\begin{deluxetable}{lrrrr}
\tablenum{2}\label{Tab:finalmodels}
\tablewidth{0pt}
\tablecaption{Changes in Parameters for Magnified Model Atmospheres}
\tablehead{
\colhead{Profile} & \colhead{$\Delta$\teff} & \colhead{$\Delta$\logg} & 
\colhead{$\Delta$ [Fe/H]} & \colhead{$\Delta\xi$}} 
\startdata
\multicolumn{5}{c}{Original \teff=5000K} \\
Profile 1 &  $-$4.0 &    $-$0.006 & $-$0.005 &   0.002 \\
Profile 18 &  23.5   &  0.024 &  0.008&    0.031\\
Profile 35 & $-$40.3  &  $-$0.039 & $-$0.018&   $-$0.059  \\
\multicolumn{5}{c}{Original \teff=5100K} \\
Profile 1 &  $-$4.7  &  $-$0.006 &  $-$0.005 &   0.002 \\
Profile 18 &  23.8   &  0.028 & 0.008 &    0.036\\
Profile 35 &  $-$31.9 &   $-$0.010 & $-$0.015&   $-$0.047\\
\multicolumn{5}{c}{Original \teff=5200K} \\
Profile 1 & $-$5.6  &  $-$0.005 & $-$0.001 &   $-$0.013\\
Profile 18 & 21.7   &  0.025  &  0.007&   0.035\\
Profile 35 & $-$39.6 &   $-$0.046 & $-$0.008 &   $-$0.048\\
\multicolumn{5}{c}{Original \teff=5300K} \\
Profile 1 & $-$4.7 &   $-$0.005 & $-$0.001 &   $-$0.013\\
Profile 18 &  20.6 &    0.025 &  0.008&    0.032\\
Profile 35 & $-$40.2 &    $-$0.049 & $-$0.018&   $-$0.065\\
\multicolumn{5}{c}{Original \teff=5400K} \\
Profile 1 & $-$4.5  &  $-$0.007 & $-$0.003&   $-$0.010\\
Profile 18 & 20.8  &   0.022 & 0.005&    0.041\\
Profile 35 & $-$36.9 &   $-$0.055  & $-$0.017&  $-$0.059 \\
\multicolumn{5}{c}{Original \teff=5500K} \\
Profile 1 & $-$4.4  &  $-$0.016 & $-$0.001 &   $-$0.003\\
Profile 18 & 19.8   &  0.020  &  0.007&   0.039\\
Profile 35 &  $-$34.5 &   $-$0.060 & $-$0.016&   $-$0.062\\
\multicolumn{5}{c}{Original \teff=5600K} \\
Profile 1 &  $-$4.2  &  $-$0.007 & $-$0.004&   $-$0.002\\
Profile 18 &  21.2   &  0.026 & 0.008&    0.041\\
Profile 35 & $-$35.7 &   $-$0.052 & $-$0.020&   $-$0.063\\
\multicolumn{5}{c}{Original \teff=5700K} \\
Profile 1 &  $-$5.0  &   $-$0.003 & $-$0.003&   $-$0.010\\
Profile 18 & 20.1    & 0.026  & 0.006 &   0.045\\
Profile 35 & $-$38.2 &   $-$0.051 &$-$0.017 &   $-$0.075\\
\multicolumn{5}{c}{Original \teff=5800K} \\
Profile 1 & $-$6.8  &  0.015 & $-$0.003&   $-$0.018 \\
Profile 18 &  25.7  &   0.023  & 0.010&   0.054\\
Profile 35 & $-$36.9 &   $-$0.047 & $-$0.019&   $-$0.079\\
\multicolumn{5}{c}{Original \teff=5900K} \\
Profile 1 & $-$14.4   & $-$0.043 & $-$0.007&    $-$0.013\\
Profile 18 & 15.9    & 0.002  & 0.005&   0.044\\
Profile 35 & $-$45.6 &   $-$0.094 & $-$0.023&   $-$0.073\\
\multicolumn{5}{c}{Original \teff=6000K, high gravity} \\
Profile 1 & $-$4.0   & $-$0.033 & 0.001&   $-$0.011\\
Profile 18 &  20.7   &  0.019 & 0.008&    0.045\\
Profile 35 & $-$35.8 &   $-$0.059 & $-$0.014&   $-$0.071\\
\multicolumn{5}{c}{Original \teff=6000K, low gravity} \\
Profile 1 & -6.9  &  $-$0.033 & $-$0.002&   $-$0.003 \\
Profile 18 & 18.8 &    0.026 & 0.004&    0.042\\
Profile 35 & $-$39.6&    $-$0.063 & $-$0.015&   $-$0.072\\
\multicolumn{5}{c}{Original \teff=6100K} \\
Profile 1 &  $-$7.0 &   $-$0.010 & $-$0.003&   $-$0.010\\
Profile 18 & 19.7   &  0.052  &0.004 &   0.031\\
Profile 35 &  $-$40.1&    $-$0.063 &$-$0.019 &    $-$0.082\\
\multicolumn{5}{c}{Original \teff=6200K} \\
Profile 1 & $-$4.9 &   $-$0.020& 0.000 & $-$0.014 \\
Profile 18 & 18.1  &   0.033  & 0.007&   0.034\\
Profile 35 &  $-$39.2 &   $-$0.096 & $-$0.019 &   $-$0.073\\

\enddata
\end{deluxetable}

\section{Application to the Study of Microlensed Bulge Dwarfs}

As stated in the Introduction, the emphasis on high magnification events 
means that in many cases there will be DLM when the spectra are taken. Our
ability to determine the effect on the spectra from observations falls into
four different possibilities: 1) the lightcurve
is indistinguishable from a point source and DLM is truly negligible;
2) the lightcurve is indistinguishable from a point, but DLM occurs;
3) the lightcurve shows extended source effects, which also means
that DLM occurs; 4) there is no photometric data for crucial parts
of the event. Of these possibilities, the first means that the
spectrum can be analyzed using standard spectroscopic techniques, while
the third can either be analyzed with standard techniques, leading
to errors of the order presented here, or analyzed using spectra with the
DLM derived from the lightcurve. It is cases 2) and 4) that are
of the most concern for ensuring reliable abundance results.
Case 2) is fully addressed by Event A, which illustrates the maximum
DLM that is possible without giving rise to noticeable effects,
given a well-covered light curve.  Case 4), i.e., poor light curve
coverage, must be handled on a case-by-case basis.  To date, all
microlensing events with spectra have had adequate photometric
coverage.

\section{Conclusion}

Microlensing of bulge dwarfs provides the exciting 
opportunity to obtain high S/N and high-resolution spectra of stars that
otherwise would be unfeasible with current telescopes. However, microlensing
can produce finite source effects that affect the spectra by 
differentially magnifying annuli in the star compared to an unmagnified
star. We have examined the practical impact of such
DLM by synthesizing spectra of dwarfs with magnification profiles
similar to events that have been observed and performing an abundance
analysis. We find that, as expected, there are changes to the
spectrum, which results in changes in the atmospheric parameters, 
metallicities and abundance ratios. Given the small
size of the changes, the S/N of the observed spectra and
the accompanying error in the atmospheric parameters, and the
length of the exposures compared to the duration of a specific
magnification profile, the resulting effect on the abundances is
small compared to other sources of error. We note that the change
in abundances can be either positive or negative (see Profile 35 vs. 
Profile 18) depending on the form of the magnification profile. Microlensing
events are more commonly in the shape of Event A, with mild magnification
causing limb brightening, so there will be a tendency for events to
be biased to lower metallicity,but Figure 11 shows that Event A leads
to changes of $<0.01$ dex in metallicity. These results,
combined with the fact that finite source effects do not affect all of the
 high-magnification cases for which dwarfs have been observed,
demonstrate that this is not the cause of any differences between
the observed metallicity distribution function for giants and dwarfs.
Finally, if the lightcurve shows even more extreme finite
source effects than have been modeled in this paper, magnification
profiles for that event can be constructed and model spectra calculated
that appropriately take those profiles into account, so this will
not be a limiting problem for the accuracy of abundances derived
for microlensed dwarfs.

\acknowledgments

We thank Carlos Allende Prieto and Judy Cohen for useful comments
and discussion. Thomas Masseron provided vital help in the setup
and use of Turbospectrum.

\end{document}